\documentclass[letterpaper, 10 pt, conference]{ieeeconf}  % Comment this line out if you need a4paper

\IEEEoverridecommandlockouts                              % This command is only needed if 
\usepackage{hyperref}
\usepackage{amsmath}
\usepackage{graphicx}
\usepackage[]{algorithm2e}
\usepackage{xcolor}

\title{\LARGE \bf
Learning Precisely Timed Feedforward Control\\ of the Sensor-Denied Inverted Pendulum\vspace{-.1in}
}

\author{Thomas L. Mohren$^{1,*}$, Thomas L. Daniel$^{2}$, Steven L. Brunton$^{1}$% <-this % stops a space
\thanks{$^{1}$Thomas L. Mohren and Steven L. Brunton are with the Department of Mechanical Engineering, University of Washington, Seattle, WA 98195, USA. 
        {\tt\small email: $^*$tlmohren@uw.edu; sbrunton@uw.edu}  *Corresponding Author}%
\thanks{$^{2}$Thomas L Daniel is with the Department of Biology, University of Washington, Seattle, WA 98195, USA
        {\tt\small danielt@uw.edu}}   
\vspace{-.1in}}

\usepackage[switch]{lineno}
\usepackage{soul}
\begin{document}

\maketitle
\thispagestyle{empty}
\pagestyle{empty}
%----------------------------------------------------------------------------------------
%Abstract------------------------------------------------------------------------------
%----------------------------------------------------------------------------------------
\begin{abstract}  
Time delays due to signal latency, computational complexity, and sensor-denied environments, pose a critical challenge in  both engineered and   biological control systems.  
In this work, we investigate biologically inspired strategies to develop precisely timed feedforward control laws for engineered systems with large time delays.  % remove learning in strategies? 
We demonstrate this approach on the nonlinear pendulum with partially denied observations, so that it is only possible to measure the state of the system near the upright position.  
Given a large disturbance that overwhelms the local feedback controller, it is necessary to add or remove energy from the pendulum so that it returns to the upright position after one full revolution.  
The partial observation near the upright position introduces a significant delay between observations and the region where actuation is most effective.  
Thus, we develop a learning algorithm that integrates sensor information into a precisely timed feedforward control signal to overcome this delay with minimal computation, training data, and set of control decisions.  
This simple controller can serve as a model for many biological systems, and can be implemented in engineered systems with time delays. 
% KEY LANGUAGE TO SAY THIS
% precisely timed, triggered, delayed feedforward control law
% sensor-denied environments; 
% novelty: we introduce a new benchmark problem
% partially-sensor blind
\end{abstract} 
% (not exceeding 200 words)

\vspace{0.5em} 
\noindent\textbf{\small Index terms: Biologically-inspired methods, Machine learning, Timing-based control, Denied measurements, Nonlinear control} 
%Biologically-inspired methods
%Machine learning
%Optimization
%\textred{new text} 
%\textblue{terms uncertain} 
% 
%\textviolet{Tom comments}
%
%\url{http://ieee-cssletters.dei.unipd.it/Page_specialissue.php}
%
%
%\textviolet{as a general rule, I prefer "engineered systems" to "engineering systems"....}

%\st{ strike out }
%-----------------------------------------------------------------------------------------
%Introduction -------------------------------------------------------------------------
%-----------------------------------------------------------------------------------------
\section{INTRODUCTION} 
 
% paragraph 1 one paragraph summary/inspiration-------------------------------------------------------------------------
Time delays between sensor measurements and control actions pose a significant challenge in engineered control systems, degrading robust performance and eventually leading to instability in the closed-loop system~\cite{Doyle2013book}.  
There are numerous sources of delay, including in the sensor and actuator hardware, in signal transduction, and in the computation of a control action.  
However, many biological systems maintain robust control performance despite large delays in their sensorimotor control circuits, providing proof-by-existence that it is possible to effectively manage these delays.  
For example, a baseball batter will initiate the swing before the ball leaves the pitcher's hand~\cite{muller2012expert}. 
Similarly, information processing in the eyes and brain of a fly and other insects takes far longer than one of its wing strokes~\cite{land1974chasing,theobald2009wide}.
Delay, in combination with limited computation power, a complex and uncertain environment, and a large set of control actions provides a compelling and relevant set of challenges for modern engineered control systems.  
Even when the control system has low latency, partially denied sensor environments will lead to delays between sensing and actuation. 
Many biological control systems employ a strategy of fast feedforward control in sensor-denied environments~\cite{cheng2018reaching} that may be used in concert with a slow supervisory feedback control to achieve the incredible observed performance and robustness~{\cite{cowan2014feedback}}. 
In this work, we will investigate biologically-inspired control strategies to learn precisely-timed feedforward control actions that overcome time delays arising from a partially denied sensor.   Event-based feedforward control with delays requires learning in both engineered and natural systems.  In biology, motor babbling in infants underlies much of the learned feedforward controls that eventually develop~\cite{meltzoff1997explaining}.  
 In engineered systems, such motor babbling is data-intensive, involving either large simulations from which actions are learned or a physical instantiation in robotic devices~\cite{takahashi2015effective}. 
%  \url{https://ieeexplore.ieee.org/stamp/stamp.jsp?tp=&arnumber=7353750     }

There are considerable current research efforts to understand and distill how biological systems handle {perturbations and} large time delays in their control architectures~{\cite{full1999templates}}. 
In the nervous system of animals, information is conveyed through  neurons by means of {discrete} action potentials.
The timing of these discrete events conveys information and greatly affects muscle activation~\cite{mohren2019coriolis,sponberg2012abdicating}.  
{Furthermore, event-triggered sensing and control are known to have advantages in energetic and computational cost compared to common control architectures in engineered systems~\cite{heemels2012introduction}.}
% do we talk about mylenation and slow transduction in invertebrates 
%Aspects of control in living systems are decentralized. 
In biology, event-based sensing is exceedingly common, to the extent that computation partially takes place at the sensor level~\cite{mohren2018neural}.  Indeed, action potentials represent timing events that are computationally efficient and reduce noise~\cite{sengupta2010action}. 
%\url{  https://journals.plos.org/ploscompbiol/article?id=10.1371/journal.pcbi.1000840}
%\st{Even though individual sensors and systems might be inaccurate, the overall system does not need to be~\cite{nakahira2019diversity}.
%Feedforward control is used by dragonflies to capture fast prey~\cite{mischiati2015internal}, and is common in human control as well~\cite{bastian2006learning}.  }

% delays can be advantageous ---------------------------

The timing-based feedforward control strategy observed in biological systems, such as prey capture by dragonflies~\cite{mischiati2015internal} or human motor contro~ \cite{bastian2006learning}, motivates an investigation to determine if there are advantages to this approach in engineered systems, or if this is simply an idiosyncrasy of the biological hardware.  
In some situations, time delays have been shown to have advantages in control design, such as deadbeat control of continuous systems~\cite{watanabe1996recent}, stabilizing oscillatory systems~\cite{abdallah1993delayed}, and in simplifying control design~\cite{lavaei2010simple}. 
We formulate the timing-based feedforward control design as an optimization problem, as is standard in control theory, and leverage the wide range of powerful optimization techniques~\cite{dp:book}.  
If this optimization is performed online from experiential data, then we may call this a \emph{learning strategy}.  
Indeed, techniques in machine learning are being rapidly integrated into control design~\cite{Brunton2019book}, including for model predictive control with deep neural networks~\cite{lenz2015deepmpc,Kaiser2018prsa,baumeister2018deep}, reinforcement learning~\cite{Sutton1998book,mnih2015human,lillicrap2015continuous}, genetic programming control in fluids~\cite{Duriez2016book,Brunton2015amr,Brunton2020arfm}, and iterative learning control.  
A number of compelling examples have explored optimizations to learn biologically inspired maneuvers related to flight~\cite{Kim2004nips,Tedrake2009isrr,novati2019controlled} and swimming~\cite{verma2018efficient}. 
Many of these learning approaches may be used to learn the optimal timing-based \emph{triggered} feedforward control explored here.  
%\textred{Reference on denied environments~\cite{arvelo2018optimal}.} 

\begin{figure}[t!]
\centering  
\hspace{-.1in}
\includegraphics{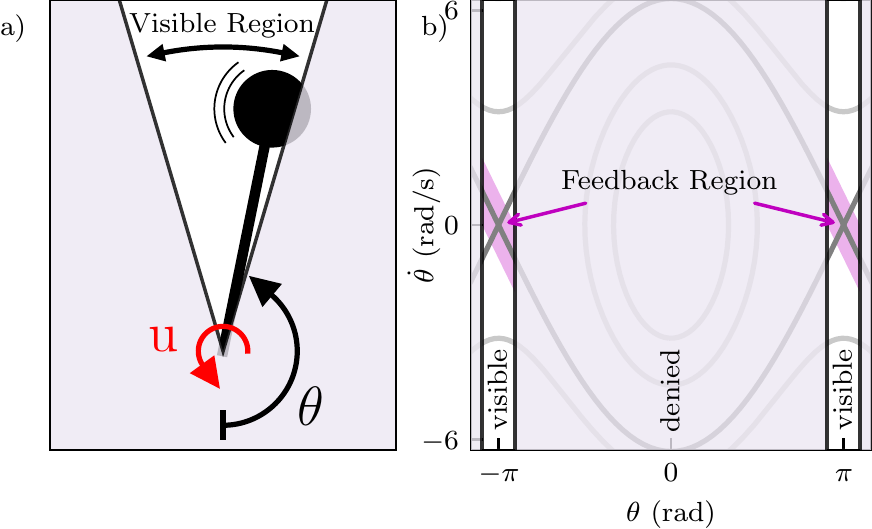} 
\vspace{-.1in}
 \caption{The pendulum model with denied sensory information, except in a small region near the inverted equilibrium. State measurements are available only in the region $\pi \pm 0.3$ radians. In the visible region, the pendulum can be controlled with state feedback. In the rest of the state space, measurements are denied. Open loop control torque is still possible in the denied region.  }
 \vspace{-.2in}
\label{fig:fig2}
\end{figure} 

% paragraph 2 relevant literature  -------------------------------------------------------------------------
To demonstrate this timing-based control strategy, we will investigate the simple pendulum, a common and well-studied benchmark for non-linear control~\cite{boubaker2012inverted}.  
There are several approaches to designing controllers for such nonlinear systems, such as energy methods~\cite{aastrom2000swinging}, relying on passivity properties of the system~\cite{lozano1998passivity}, and LQR-trees~\cite{tedrake2010lqr}. 
The controller typically has a hybrid form, consisting of a nonlinear controller for the swing up and a linearized feedback controller to stabilize the inverted equilibrium~\cite{boubaker2012inverted}. 
In the case of limited or saturated actuation, bang-bang control generally provides an optimal minimum-time controller~\cite{furuta1991swing,boubaker2012inverted}. 
Discrete timing control has also been used on the pendulum~\cite{bhounsule2015discrete}.

%Unlike these previous studies, we will consider the pendulum with partially denied observations, so that it is only possible to measure the state of the system near the upright position where LQR is effective, as shown in Fig.~\ref{fig:fig2}.  
%When the system experiences a large disturbance and leaves the LQR region of convergence, it is necessary to add or remove energy so that it returns to the upright position after one full revolution.  

\begin{figure}[t]
%\vspace{-.1in}
\centering 
\includegraphics{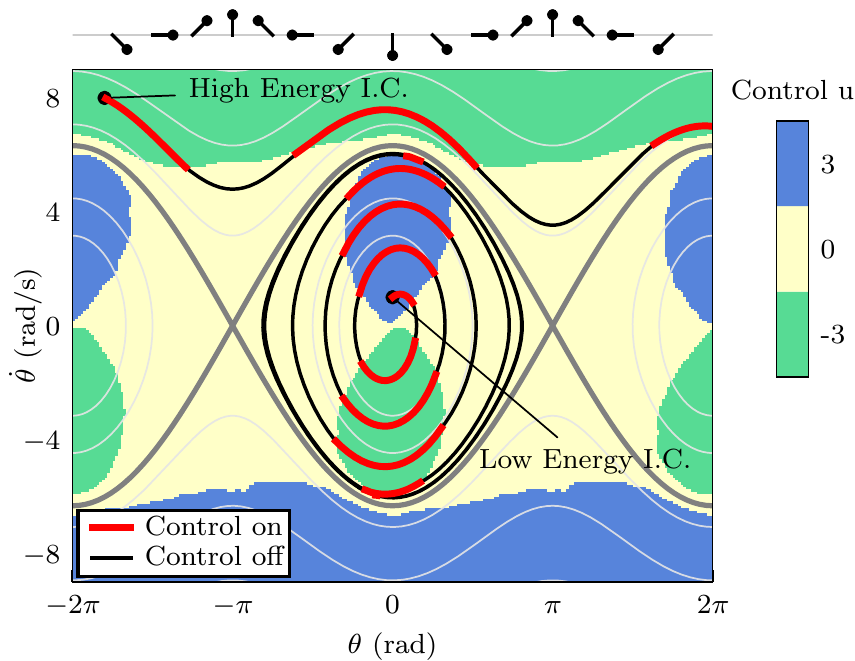} 
\vspace{-.25in}
 \caption{The state space diagram of the optimal controller for the simple pendulum with bang-bang control and cost ratio $R/Q = 100$, where R is control cost, and Q is the cost of error in the pendulum energy. 
 Two initial conditions (I.C.s) are shown, one with a low energy starting point, requiring pumping up to obtain the inverted equilibrium, and one high energy starting point, requiring energy bleeding at certain phases of the state space. 
 Blue and green regions of the state space are regions where control is on, as denoted by the red colored sections of the trajectories. These regions of effective actuation will be learned from an online optimization based entirely on information in the visible region in Fig.~\ref{fig:fig2}.}
\label{fig:fig1}
\vspace{-.2in}
\end{figure}

Unlike these previous studies, we will consider the pendulum with partially denied observations (partially denied sensing), so that it is only possible to measure the state of the system near the upright position where  feedback  is effective, as shown in Fig.~\ref{fig:fig2}.  
When the system experiences a large disturbance and leaves the feedback region of convergence, it is necessary to add or remove energy so that it returns to the upright position after one full revolution.  
The optimal time to pump in energy is at the point of maximum velocity (shown in Fig.~\ref{fig:fig1}), as in the case of a child on a swing~\cite{piccoli2005pumping}. 
However, the system is not observable in this region of space, and so the proper control action must be pre-planned and timed based on the sensor information near the upright position.  
Thus, partial observation near the upright position introduces a large delay between observations and the region where actuation is most effective, making this a suitable problem to explore our timing-based feedforward learning strategy.
In particular, we leverage optimization based on limited measurement data from the pendulum to \emph{learn} the appropriate control action and timing, demonstrating the ability to overcome the gap between information availability and control effectiveness.

\section{BENCHMARK PROBLEM}

% Pendulum model-----------------------------------------------------------------------------
Our benchmark system is the simple frictionless pendulum with a single degree of freedom $\theta$ and a torque input $u$. 
With a massless rod, the pendulum equation of motion is 
 \begin{align} 
ml^2 \ddot{\theta} & = mgl \sin \theta  + u,
 \end{align}
which may be written in terms of $x_1=\theta$ and $x_2=\dot{\theta}$
 \begin{align} 
\dot{x}_1 & = x_2\\
\dot{x}_2 & = \frac{g}{l}\sin(x_1) + \frac{1}{ml^2}u.
 \end{align}  
We choose non-dimensional parameters $l=1\,$m, $m=1\,$kg,  $g=-10\,$N/(ms$^2$). 
The controller saturates at a torque of $3\,$Nm, which is the torque due to gravity at a pendulum angle of approximately $0.3\,$rad ($17\,$degrees).

\begin{figure*}[t]
\centering 
\includegraphics{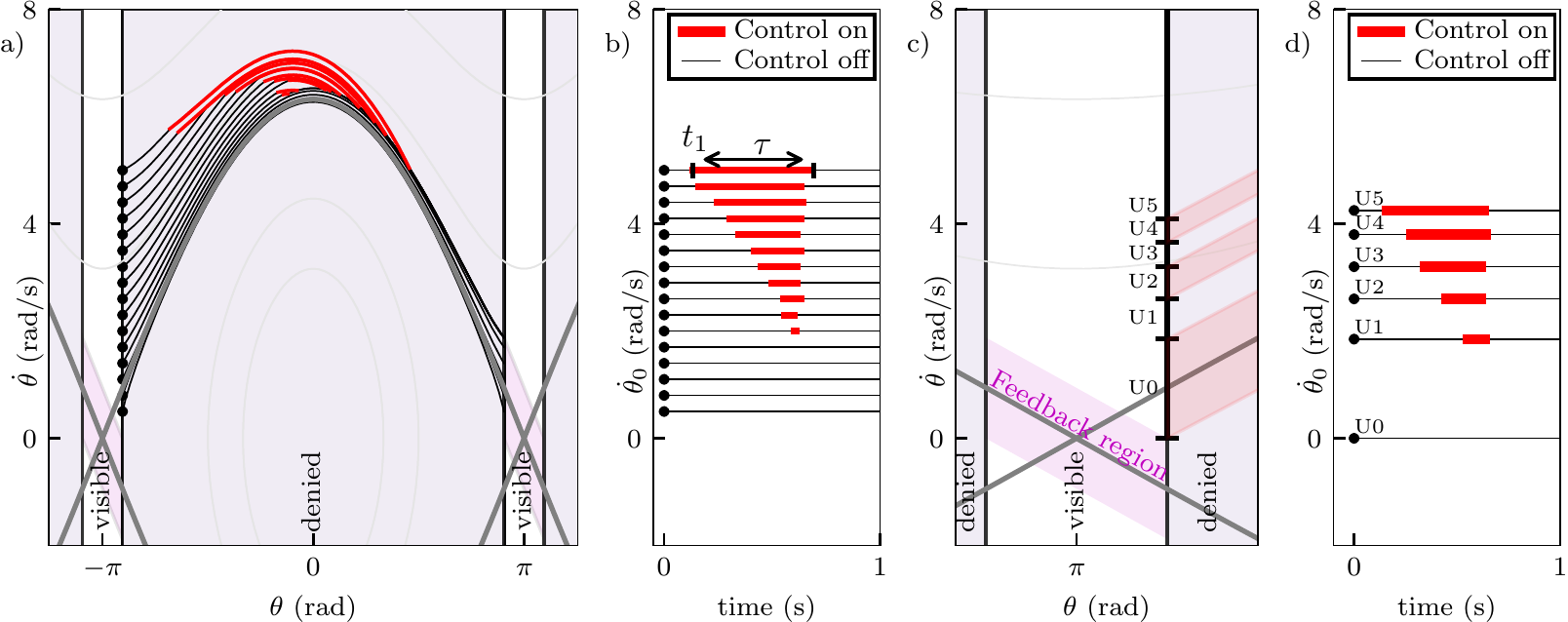}  
\vspace{-.1in}
 \caption{ (a) For an array of initial velocities $\dot{\theta}_0$, we optimize the start-time $t_1$ and duration $\tau$ of the feedforward, bang-bang control protocols to drive the pendulum to the feedback attracting region after one revolution.
(b) For low initial velocities $\dot{\theta}_0 \le 1.8$, no control is required to swing back to the attractor. 
For increasing velocities, the start-time gets smaller, and the duration gets longer. 
(c) We find a subset of feedforward trajectories that converge into the feedback attractor. 
(d)  By deciding on one out of these 6 protocols, any initial velocity $0\geq \dot{\theta}_0 \leq 5$ will be returned to the feedback region. }
\label{fig:fig3}
\vspace{-.1in}
\end{figure*}

In our example, full observations of the state (i.e., pendulum angle and angular velocity) will be available in a limited region near the upright position, within the range $  \pi - 0.3 \geq \theta \leq \pi + 0.3 $. 
%This leaves a narrow region that can be controlled by full state feedback via a linear quadratic regulator (LQR). 
This leaves a narrow region that can be controlled by full state feedback. %\st{via a linear quadratic regulator (LQR)}. 
However, if a disturbance drives the pendulum outside of this visible region, the controller is sensor-blind and must use limited information from the visible region to pre-plan a precisely timed feedforward control action.  
This obscured state space is shown in Fig.~\ref{fig:fig2}. 

Next, we will describe the full-state feedback controller near the pendulum-up configuration and the global cost function that is used to optimize the feedforward control in the sensor-denied region.  
The goal will be to learned the regions of effective actuation in Fig.~\ref{fig:fig1}, and the precise time to trigger actuation, from an online optimization based entirely on information in the visible region in Fig.~\ref{fig:fig2}.

%% homoclinic orbit description ------------------------------------------------------------------------------
%Because there is no damping and no disturbance in the system, the pendulum's total energy is constant without forcing. 
%The resulting energy levels form orbits in state space that the pendulum will follow. 
%The most notable is the homoclinic orbit, which has the same total energy as the inverted equilibrium. 
%When the pendulum is on this orbit, it will move into the inverted equilibrium over time. 
%The equation for the homoclinic orbit is: 
%\begin{align}
% \dot{\theta}  & =  \pm \sqrt{  \frac{-2g }{l} } \sqrt{1+\cos \theta} \text{ , with derivative: }  \\ 
% \frac{ \partial \dot{\theta} }{ \partial \theta} & = \pm \sqrt{  \frac{-g }{ 2 l} }  \frac{ \sin \theta} {  \sqrt{1+\cos \theta}  } 
% \end{align}

% Feedback control -----------------------------------------------------------------------
\subsection{Feedback control near the inverted fixed point}
The pendulum has two fixed points, one in the neutrally stable downward position, and another the unstable inverted position. 
After linearization about the inverted equilibrium $x_1=\pi$, so that  $\mathbf{x} = \begin{bmatrix} \theta - \pi , \dot{\theta} \end{bmatrix}^T$, the pendulum can be made stable by applying full-state feedback control $u=-\mathbf{K}\mathbf{x}$:
\begin{align*}
\dot{\mathbf{x}} = (\mathbf{A}- \mathbf{BK} )\mathbf{x} .
\end{align*}
 
We choose gains such that the eigenvector of the state feedback system aligns with the homoclinic orbit connecting the unstable saddle at the inverted equilibrium to itself through one revolution of the pendulum:
\begin{align*}
\frac{x_2}{x_1}& =  \frac{ \partial \dot{\theta} }{ \partial \theta} \bigg|_{\pi^-}  =  -\sqrt{  \frac{- g }{l} }.
\end{align*}
We then solve for the gains in $\mathbf{K}$, noting that
\begin{align*}
(\mathbf{A}-\mathbf{B K} ) \mathbf{x} &= \lambda \mathbf{x } \quad\Longrightarrow\quad k_2      =   k_1   \sqrt{  \frac{- g }{l} }  +2    ml^2  \sqrt{ \frac{ -g }{l }  }.   
 \end{align*}
 Given a choice of $ k_1 = 1000$, this results in a gain $K=[1000,309.9]$ with saturation at $u =3$Nm. 
This gain creates the parallelogram-shaped region of attraction near the inverted equilibrium, referred to as the Feedback region in Fig.~\ref{fig:fig3}c.

 \subsection{Optimal control}
% defining optimal controller-----------------------------------------------------------------------
%Feedback control with the saturated controller is only effective near the inverted equilibrium. 
Outside the feedback controller region of attraction, we aim to find a globally optimal controller that returns any initial condition to the inverted equilibrium within a reasonable amount of time and without expending too much actuation energy. 
We define our cost function to be a cumulative penalty on a combination of state error and control effort.  
Instead of a standard state error term given by the Euclidean distance to the goal state, we instead use the difference between the total energy of the system and the goal energy at the inverted equilibrium. 
The cost is then given by:
 \begin{align}
 J & = \int_0^\infty\left( Q  e^2 + R u^2\right)\,d\tau,
%J & = \Delta t \sum_{i = 1}^n Q e_i^2  + R u_i^2, 
 \end{align}
with $e = E_\text{total} - E_\text{final}$, $Q=1$, and $R=100$.  
Outside of the feedback region we use a bang-bang controller with $u \in \{-3,0,3\}$Nm.

Before describing the results of our procedure to learn the optimal control given only information in the sensor-visible region, here we derive the global optimal controller with full information via dynamic programming~\cite{bertsekas1996neuro}, shown in Fig.~\ref{fig:fig1}.   
This optimal bang-bang controller will provide a benchmark for our learning results in the next section.  
%
%% GOOD INFO HERE... LETS SAVE FOR REVISION
%We implement this by having an array of initial conditions distributed over the homoclinic orbit. 
%All points take three different actions, according to the discrete control actions $u= \{ -3,0,3\}$, with randomized timestep $\Delta t = 0.025 + R$, where $R$ is a random number between $\left[ 0, 0.05 \right]$ drawn from a uniform distribution.  
%The randomness in time step reduces periodic noise in the solution. 
%For all new points, they are compared to all points within a 0.02 radius in state space. 
%If they are the lowest cost J within in that space, they survive and other points (if present) with a higher cost in that circle are forgotten.  
%We continue time stepping until no new point survives, and thus all of the reachable state space is explored. 
%This provides a cost map over the entire state space. 
%We then apply a Gaussian filter to obtain a smooth cost function over the space. 
%With this cost function, we can evaluate the best control action at any location.
%The described method gives us the control map in figure \ref{fig:fig1}.  
%
% discussion of optimal control result--------------------------------
%Unlike minimum-time control, the torque is not always active with t. 
With the prescribed cost function, control is active only during phases of the swing where velocity is highest, and thus the most change in energy is achieved: 
\begin{align}
\Delta E  & = \int_{t_1}^{t_1+ \Delta t} u \dot{\theta}(\tau)\,   d\tau.  \label{eq:energy}
%I & = \int T \dot{ \theta} (t) dt 
\end{align} 
From a low-energy initial condition, it will take multiple swings to pump up to the homoclinic orbit (Fig.~\ref{fig:fig1}). 

 \begin{figure*}[t ]
\centering 
\includegraphics{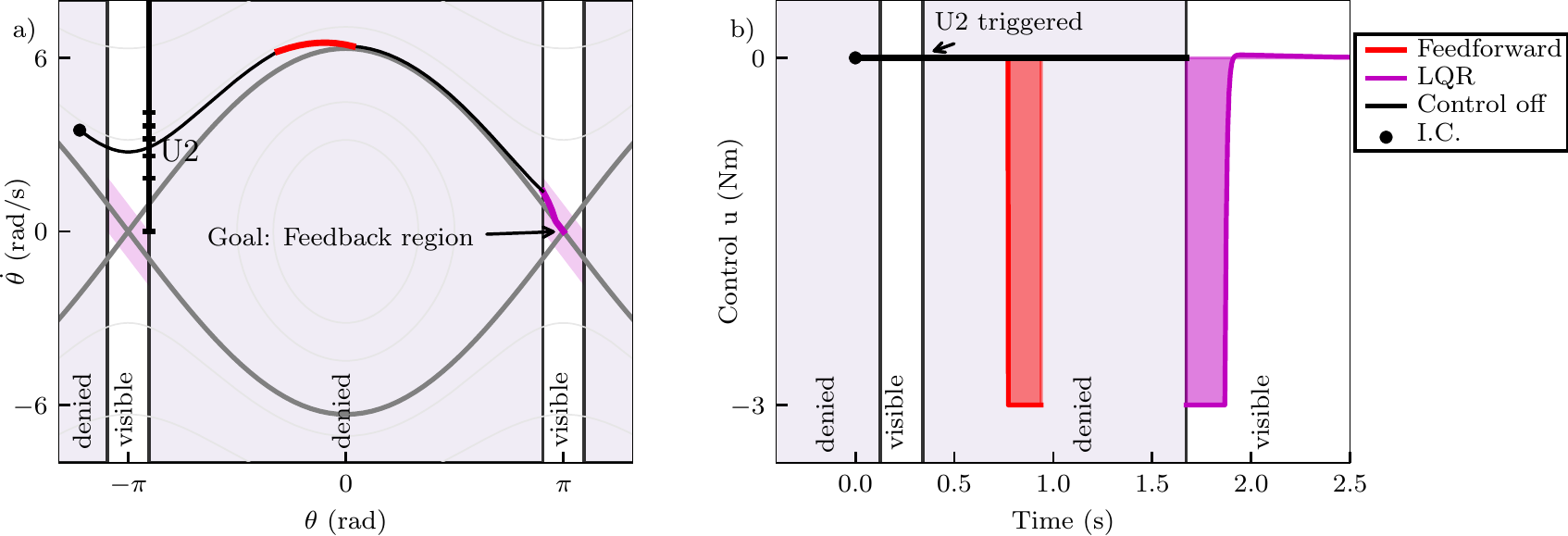} 
\vspace{-.1in}
 \caption{  
The combination of feedforward and feedback controller implemented on the partially observed pendulum. 
The pendulum first moves through the visible region, and as it enters the sensor-denied environment, the $U_2$ control protocol is selected. 
After time $t_1$ the control is triggered, remaining on (red) between $t_1 \leq t \leq  t_1 + \tau$. 
The protocol is such that, once the pendulum reaches the visible region again, it now falls in the feedback region of attraction (purple).
Figure (a) shows the trajectory in state space, (b) shows the timing and magnitude of the controller. 
    }
\label{fig:fig4}
\vspace{-.1in}
\end{figure*}

\section{LEARNING TIMING-TRIGGERED CONTROL}
% Feedforward controller method---------------------------------------------
Now we consider the problem of learning a near-optimal feedforward control law when it is impossible to measure the state when the control is most effective. 
%In particular, we will consider trajectories originating in the sensor-visible window that have been pushed outside of the LQR effective region by a large disturbance.  
In particular, we will consider trajectories originating in the sensor-visible window that have been pushed outside of the feedback effective region by a large disturbance.  
This controller will necessarily need to overcome the time-delay between sensing and effective actuation, which will manifest as a pre-planned control strategy consisting of bang-bang control at a precise future time. 
The decision on what feedforward protocol to use is made on the decision boundary, which we define to be the boundary in state space where the pendulum leaves the visible region. 
The feedforward controller is parameterized by the time $t_1$(s) from the decision boundary that the bang-bang control is turned on, and the duration of control $\tau$(s). 
Note that the optimal control parameters will be a function of the energy of the pendulum as it leaves the visible region, as shown in Fig.~\ref{fig:fig3}.  
The control parameters are optimized to minimize the cost function from above, with an additional penalty for failing to reach the feedback region of attraction at the end of one revolution.  
First, we will optimize the control parameters in an \emph{offline} learning procedure to demonstrate the process.  Afterward, we will describe an \emph{online} exploration and optimization procedure that provides a more realistic and useful learning strategy.

% Feedforward controller implementation ---------------------------------------
%We can now find the parameters for the optimal feedforward controller.
% Result 1  timing-based feedforward solves obscured state ---------------------
Starting from an array of initial conditions on the decision boundary, we find the associated set of optimal $t_1$ and $\tau$ parameters  (Fig.~\ref{fig:fig3}b). 
The cost function is not differentiable over $t_1$ and $\tau$, and there is a large ``valley" of near-optimal parameter combinations (see appendix for details).  Therefore, we use a Nelder-Mead optimization, in combination with basin-hopping to escape potential local minima. 
Obtaining start-times and durations for various trajectories leaving the visible region, we observe a near-linear increase in duration. 
As the initial velocity increases, the total energy of the pendulum increases quadratically, whereas total change in energy achieved over the same time increases linearly with velocity (Eq.~\ref{eq:energy}), thus requiring a longer control duration. 
We can attribute the earlier start-times $t_1$ with increasing initial velocity $\dot{\theta}_0$ to two causes: 
1) the entire trajectory is faster, so the apex of the trajectory occurs earlier, and 2), the longer duration control-on phase has to be centered around the apex, thus leading to an earlier start-time. 
Furthermore, we note that trajectories with low initial velocity ($\dot{\theta}_0 \le 1.8$ rad/s) reach the feedback region without requiring the controller to turn on during the obscured region (Fig.~\ref{fig:fig3}b). 
%Interestingly, within the visible region there are parts of state space where leaving the visible region, having no feedforward protocol, and enacting feedback control once back in the visible region, is the only way to obtain the inverted equilibrium.  
Interestingly, there are parts of the visible state space outside of the feedback region, where the only way to reach the feedback region is to swing through without any control. 
For higher initial velocities  $\dot{\theta}_0 > 6.1$ rad/s (not shown in Fig.~\ref{fig:fig3}), the feedforward controller will be on at all times, while still not reaching the inverted equilibrium.

% Result 2 grouping of controller, only 6 gruops required---------------------------------------------------------------------
Next, we seek to reduce this set of feedforward controllers into a minimal set, which may be interpreted as a decision on the visible boundary about which feedforward control protocol to trigger (see Fig.~\ref{fig:fig3}c-d).  
In the biological setting, it is feasible that there may be advantages to having a smaller set of controllers to choose from.  
We determine this minimal set by starting with the lowest initial condition protocol, and increasing the initial velocity until the control protocol no longer drives the trajectory to the feedback region within a single revolution of the pendulum. 
At that limiting initial velocity, we then find the control protocol with the highest initial velocity that still drives the pendulum to the feedback region within one revolution.  
This procedure is continued until we obtain a minimal set of 6 control protocols that will drive initial velocities $ 0 \leq \dot{\theta}_0\leq 5$ to the feedback region (Fig.~\ref{fig:fig3}d).  
We observe that for lower velocities, no action is required to return to the inverted equilibrium.
For higher velocities, the duration increases and the start time decreases, which is to be expected since the protocols are drawn from the larger set of protocols in Fig.~\ref{fig:fig3}. 
We note that for higher initial velocities the protocols are spaced closer together. 
This is again due to the kinetic energy increasing quadratically.

% Hybrid set implemented ----------------------------------
 % Result 3, implementation of hybrid  -----------------------------------------------------------
We now implement the small subset of triggered feedforward protocols in combination with feedback control in the region of attraction according to algorithm \ref{al:al1}. 
When the trajectory crosses the decision boundary a feedforward control protocol is triggered with the parameters $t_1$ and $\tau$.  
After time $t_1$ the bang-bang control is triggered for duration $\tau$, after which the pendulum passively evolves to the region where feedback control is activated.  
If the system is in the visible region, but not in the feedback region, no control is exerted. 
The implementation of this combined controller is shown in Fig.~\ref{fig:fig4}.
The pendulum starts from the  initial condition in the obscured region, moves through the visible region, where it is observed but then moves too fast to be stabilized to the equilibrium. 
As it moves through the decision boundary to the obscured region, the $U_2$ protocol is triggered. 
Now out of view, the controller is active between 0.43 and 0.60 seconds and is brought back close to the homoclinic orbit. 
When back in view, the pendulum enters the feedback region and it is stabilized around the inverted equilibrium with feedback.

\begin{figure}[t ]
\centering 
\includegraphics{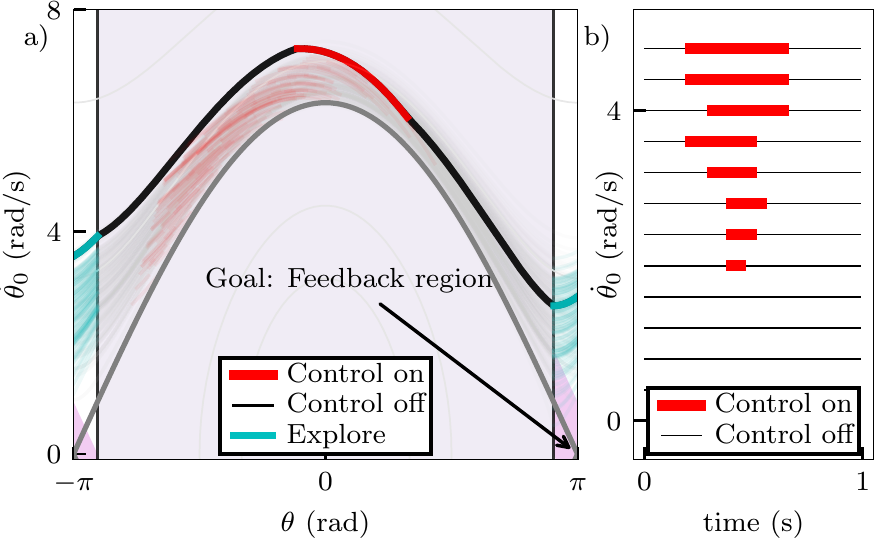} 
 \caption{ 
 We implement an online learning strategy that optimizes the parameters of the feedforward controller. 
We use an exploration protocol where the controller adds energy to the system to set up higher velocity initial conditions (a). 
Each time the pendulum leaves the visible region, a new combination of feedforward parameters $t_1$ and $\tau$ is tested. 
Using this continuous exploration and optimization, we find a set of controllers that approximates the optimal control solution (b).
 }
\label{fig:fig5}
\end{figure}

%  continuous exploration ------------------------------------------------------------------
% Result 4 continuous exploration ------------------------------------------------------------------
Lastly, we continuously explore the state space in an \emph{online} learning procedure, without resetting the simulation between initial conditions (Fig.\ref{fig:fig5}a). 
Because the controller removes energy from the system, we explore new initial conditions by adding energy in the visible region after each revolution. 
When the pendulum leaves the visible region, we try a set of $t_1$ and $\tau$, and evaluate the cost of the trial once it reaches the visible region again.  
We discretize the cost matrix as follows:  initial velocities $\dot{\theta}_0$ in bins with a 0.4 rad/s width in the range $\left[0,5\right]$,  $\tau$ in steps of 0.05 seconds in the range $\left[ 0, 0.6\right]$ seconds, and $t_1$ in steps of 0.05 in the range $\left[ 0.2, 0.6\right]$ seconds. 
To prevent the controller bleeding too much energy and thereby not entering the visible region again, we build up $t_1$ and $\tau$ from zero with discrete increments, starting from the lowest initial velocities. 
Once $\tau$ is sufficiently large to drive the pendulum back to the feedback region, we stop exploring larger $\tau$'s from that initial condition and explore higher initial velocities.  
This continuous exploration shows that with far fewer trials the set of protocols still approximates the optimal solution (Fig.~\ref{fig:fig4}b).  
However, the smaller number of trials comes at the cost of precision in both start-time and duration, although this may be refined with more data. 
Furthermore, by binning the initial conditions, suboptimal combinations might lead to the lowest observed cost as a result of a favorable initial condition.

\begin{algorithm}[t] 
 \eIf{ In feedback region}{
   $u = - K x $, with $|u| \leq 3$\; 
   }{
   \eIf{Crossing decision boundary}{
      Select control protocol $U_j$  \; 
      look up parameters $t_1$ and $\tau$ belonging to $U_j$ \; 
       Start timer $t$ \;  
       u = 0 \;
   }{
   \eIf{ $ t_1 \leq t \leq  (t_1+\tau) $ }{ 
   $u = -3$ \;
  } {
  $u = 0 $ \; 
  }
  }
  }
 \caption{Decision scheme for the combined feedback and feedforward controller}
 \label{al:al1} 
\end{algorithm}

%----------------------------------------------------------------
% Conclusions------------------------------------------------- 
%----------------------------------------------------------------
\section{CONCLUSIONS}

% Paragraph 1, review of results
In this work we explored how to learn a timing-based control strategy inspired by biological systems. 
In particular, we developed a hybrid feedforward and feedback controller for a nonlinear pendulum where it is only possible to measure the system near the upright configuration.  As is the case for motor babbling, we simulated a vast number of control actions and used a data-driven approach to learn the optimal set of control actions given limited sensing and delay.
We found that it is possible to optimize a precisely timed feedforward controller using only information available in the visible region.  For simplicity, we restricted ourselves to a bang-bang controller with two parameters -- the start time and duration.  
 The two-parameter solution was learned with relatively few trials, as is often the case in natural systems.
By grouping the feedforward protocols into a minimal set, we reduced the number of choices while still driving the system to the upright position. 
Finally, we have also developed a more realistic online learning procedure, consisting of both exploration and parameter learning. 

There are a number of important future avenues suggested by this work.  
First, we did not consider the effect of measurement noise or exogenous disturbances outside of the visible region.  This work is meant to be a proof of concept and to develop a benchmark problem to explore biological learning and control strategies. In the future, it will be important to explore the robustness of these strategies to noise and disturbances. 
Though the decision on control protocol was based solely on the system state when crossing the decision boundary, measurement noise could be addressed by using a state estimator for the the entire presence in the visible region.   Indeed, nonlinear neural-inspired temporal filters may be incorporated into the decision process serving that function. Such neural-inspired approaches are known to improve decisions and classifications in biological systems~\cite{mohren2018neural}.  
We can imagine a scenario where measurements are in fact not obscured, but where measurement noise is time dependent. 
Feedforward protocols can then be triggered when the certainty in the state estimate drops below some desired level.  
While many engineered systems currently rely on continuous sensory information and controllability, many biological systems have found ways to deal with time-varying observability and control authority.   
Thus, although this paper tested the timing-based feedforward concept on a simple non-linear problem with favorable passivity properties, we believe that applying these ideas to more complex systems will be a fruitful area of future work.

%
%A conclusion section is not required.Do not replicate the abstract as the conclusion. A conclusion might elaborate on the importance of the work or suggest applications and extensions. 

%%%%%%%%%%%%%%%%%%%%%%%%%%%%%%%%%%%%%%%%%%%%%%%%%%%%%%%%%%%%%%%%%%%%%%%%%%%%%%%%

%\addtolength{\textheight}{-2cm}   % This command serves to balance the column lengths

% TO remove 
%\clearpage 
%%%%%%%%%%%%%%%%%%%%%%%%%%%%%%%%%%%%%%%%%%%%%%%%%%%%%%%%%%%%%%%%%%%%%%%%%%%%%%%%\

%%%%%%%%%%%%%%%%%%%%%%%%%%%%%%%%%%%%%%%%%%%%%%%%%%%%%%%%%%%%%%%%%%%%%%%%%%%%%%%%
\section*{ACKNOWLEDGMENTS}
 
This work was supported by the Komen Endowed Chair  and support from the Washington Research Foundation to TLD, the Air Force Office of Scientific Research (AFOSR) under awards FA9550-14-1-0398 to TLD and 
 FA9550-19-1-0386 to SLB and TLD, and the Army Research Office (ARO) under awards W911NF-19-1-0045 and W911NF-17-1-0306 to SLB.   
We would also like to thank Bing Brunton for valuable discussions on sparse sensor systems.  
 
\section*{CODE AND DATA}

All code for the implementation of methods, generation of data, and reproducing the figures can be found on \url{https://github.com/tlmohren/timed_feedforward_control}.

% \section*{Style guides}
% 
% \url{http://ieee-cssletters.dei.unipd.it/Page_authors.php?p=1&mkt_tok=eyJpIjoiTURZM01ESmhNVFk1TlRrMiIsInQiOiJcL2E0Rk1hMVlQQ2RNWDBTdkpjalp1OWpuWjlIRU9OZnRIMDRWYnF2UFJYYlJcL1JhaHVGc1BnU0VcL1JJWlBzQUxwaDNGMEF4UXVHd1JuVnhDZ242NmpPNnlzTVg5M2NiOGpvakpyaXRUNXBZWFVtR3E5Sm5teDUzKzdWN21oRTJ5aCJ9} 
% 
% \url{https://www.ieee.org/content/dam/ieee-org/ieee/web/org/pubs/eic-guide.pdf} 

%%%%%%%%%%%%%%%%%%%%%%%%%%%%%%%%%%%%%%%%%%%%%%%%%%%%%%%%%%%%%%%%%%%%%%%%%%%%%%%%
 
%\bibliographystyle{acm}
\bibliographystyle{IEEEtran}
\bibliography{IEEEabrv,Neural_Inspired_Control_references} 
% \bibliography{./IEEEabrv,./IEEEexample}

%\bibliography{IEEEabrv,Neural_Inspired_Control_references}
%\bibliography{Neural_Inspired_Control_references}

%\begin{thebibliography}{99}
%
%\bibitem{c1} G. O. Young,  Synthetic structure of industrial plastics (Book style with paper title and editor),  	in Plastics, 2nd ed. vol. 3, J. Peters, Ed.  New York: McGraw-Hill, 1964, pp. 15 64.
%\bibitem{c2} W.-K. Chen, Linear Networks and Systems (Book style).	Belmont, CA: Wadsworth, 1993, pp. 123 135.
%\bibitem{c3} H. Poor, An Introduction to Signal Detection and Estimation.   New York: Springer-Verlag, 1985, ch. 4.
%\end{thebibliography}

\end{document}